\documentclass[12pt]{iopart}

\usepackage[dvips]{graphicx}
\input{epsf}

\def\h#1{$^{#1}$H}
\def\he#1{$^{#1}$He}
\def\li#1{$^{#1}$Li}
\def\be#1{$^{#1}$Be}

\newcommand{\nc}{\newcommand}
\nc{\hef}{$^4$He}
\nc{\het}{$^3$He}
\nc{\lisx}{$^6$Li}
\nc{\lisv}{$^7$Li}
\nc{\bes}{$^7$Be}
\nc{\beet}{$^8$Be}
\nc{\ben}{$^9$Be}
\nc{\hefm}{{\rm ^4He}}
\nc{\hetm}{{\rm ^3He}}
\nc{\lisxm}{{\rm ^6Li}}
\nc{\lisvm}{{\rm ^7Li}}
\nc{\besm}{{\rm ^7Be}}
\nc{\beetm}{{\rm ^8Be}}
\nc{\benm}{{\rm ^9Be}}
\nc{\xm}{$X^-$}
\nc{\xp}{$X^+$}
\nc{\xz}{$X^0$}
\nc{\bex}{(\bes\xm)}
\nc{\bexm}{(\besm X^-)}
\def\bee{\begin{equation}}
\def\ee{\end{equation}}
\newcommand{\fr}[2]{\frac{#1}{#2}}

\def\epstwo@scaling{0.48}

\def\showtwo#1#2{
  \centering
  \leavevmode
  \epsfxsize=\epstwo@scaling\linewidth
  \epsfbox{#1.eps} \hfil
  \epsfxsize=\epstwo@scaling\linewidth
  \epsfbox{#2.eps}
  }

\newcommand{\simge}{\,{}^>_{\sim}\,}
\newcommand{\simle}{\,{}^<_{\sim}\,}

\begin{document}

\title{Big Bang Nucleosynthesis and Particle Dark Matter}
\author{Karsten Jedamzik$^{\,(a)}$ and Maxim Pospelov$^{\,(b,c)}$}
\address{
$^{\,(a)}$ Laboratoire de Physique Th\'eorique et Astroparticules, UMR5207-CNRS,
Universit'e Montpellier II, F-34095 Montpellier, France,\\ 
$^{\,(b)}$ Department of Physics and Astronomy, University of Victoria, 
     Victoria, BC, V8P 1A1 Canada\\
$^{\,(c)}$ Perimeter Institute for Theoretical Physics, Waterloo,
Ontario N2J 2W9, Canada
}
\ead{\\jedamzik@lpta.univ-montp2.fr\\
pospelov@uvic.ca}
\begin{abstract}

 We review how our current understanding of the light element synthesis
during the Big Bang Nucleosynthesis era may help shed light on the 
identity of particle dark matter. 

\end{abstract}

\section{Introduction}

%In an attempt to explain 
%nuclear abundance patterns observed
%in the nearby Universe, such as the pecuilar high helium mass
%fraction $Y_p\approx 0.25$,
%initially speculative work by Alpher, Gamov, Fermi, 
%Follin, Hayashi, Herman, Turkevich,
%and others in the late 40s and throughout the
%50s on an era of
%nucleosynthesis (element formation)
%in an expanding Universe at very high 
%temperature $T\sim 10^9$K
%%adiabatically expanding to the present $T = 2.7$K Universe, 
%developed
%slowly but steadily over the coming decades into the standard model of
%Big Bang nucleosynthesis (BBN). 

In the late 40s and throughout the
50s  a number of visionary scientists including Alpher, Fermi, 
Follin, Gamow, Hayashi, Herman, and Turkevich attempted to explain 
nuclear abundance patterns observed
in the nearby Universe, such as the peculiar high helium mass
fraction $Y_p\approx 0.25$. This initially speculative work on an era of
nucleosynthesis (element formation)
in an expanding Universe at very high 
temperature $T\sim 10^9$K developed
slowly but steadily over the coming decades into what is now known
as the standard model of Big Bang nucleosynthesis (BBN).
The idea that the Universe may have
undergone a very hot and dense early phase got triggered by the observations
of Hubble in the 1920s, of the recession velocity of galaxies 
being proportional to their inferred distance from the Milky Way, which
were most elegantly explained by a Universe in expansion. 
The "expanding, hot Big Bang" idea received further support by the
observation of the cosmic microwave background radiation (CMBR)
by Penzias and Wilson in 1965, believed to be the left-over radiation of
the early Universe. Detailed observational and theoretical studies of BBN
as well as the CMBR and the Hubble flow have developed into the main pillars
on which present day cosmology rests.

BBN takes place between eras with
(CMBR) temperatures $T\simeq 3\,$MeV and
$T\simeq 10\,$keV, in the cosmic
time window $t\simeq 0.1-10^4\,$sec,
and may be characterized as a freeze-out from nuclear
statistical equilibrium of a cosmic plasma at very low 
$\sim 10^{-9}$ baryon-to-photon number ratio
(cf. Section~\ref{SBBNtheory}), conditions which are not 
encountered in stars. It produces the
bulk of \he4 and \h2 (D), as well as good fractions of \he3 and \li7 observed
in the current Universe, whereas all other elements are 
believed to be produced either by stars or cosmic rays.
In its standard version it assumes a Universe expanding according
the laws of general relativity, at a given homogeneously
distributed baryon-to-photon
ratio $\eta_b$, with only Standard Model particle degrees of freedom 
excited, with negligible $\mu_l\ll T$ lepton
chemical potentials, and in the absence of any significant perturbations from 
primordial black holes, decaying particles, etc. 
By a detailed comparison of observationally inferred abundances 
(cf. Section~\ref{observations}) with those theoretically predicted, 
fairly precise
constraints/conclusions about the cosmic conditions during the BBN era may
thus be derived. 
%BBN is very sensitive to the change in the timing of major BBN events such as 
%neutron/proton. Using this sensitivity, one can constrain, for example, 
BBN has been instrumental, for example, in constraining
the contribution of extra "degrees of freedom" excited in the early Universe
to the total energy density, such as predicted in
many models of particle physics beyond the standard model.
Such contributions may lead to
an enhanced expansion rate at $T\sim 1\,$MeV 
implying an increased \he4 mass fraction.
%Such contributions may lead to
%an enhanced expansion rate at $T\sim 1\,$MeV, with the resulting change in
%timing of freeze-out from neutron-proton interconversion as well as time
%of significant \he4 synthesis leading to an increased \he4 mass fraction.
%, and since BBN is very sensitive
%towards a change in the timing of major BBN events, such as freeze-out
%from weak equilibrium and neutron-proton ($n-p$) interconversion, it may lead
%to an increased $n/p$ ratio and \he4 mass fraction. 
%and an increased \he4 mass fraction.
It is now known that 
aside from baryons and other subdominant components not much more than
the already known relativistic degrees of freedom
(i.e. photons $\gamma$'s, electrons and positrons
$e^{\pm}$'s, and three left-handed neutrinos $\nu$'s)
could have been present during the BBN era. BBN is also capable of constraining
very sensitively any non-thermal perturbations as induced, for example,
by the residual annihilation of weak scale dark matter 
particles (Section~\ref{annihilation}), or by the decay of 
relic particles (Section~\ref{decay})
and the possible concomitant production of dark matter. Moreover, the
sheer presence of negatively charged or strongly-interacting 
weak mass-scale particles during
BBN (Section~\ref{catalysis})
may lead to dramatic shifts in yields of  light element
through the catalytic phenomena. BBN may therefore constrain properties and 
production mechanisms of dark matter particles, and this chapter aims at revealing this connection.

It is possible that the biggest contribution of BBN towards understanding
the dark matter enigma has already been made. Before the advent of precise
estimates of the fractional contribution of baryons to the present
critical density, $\Omega_b\approx (0.02273\pm 0.00062)/h^2$, 
where $h$ is the Hubble constant in units 100 km s$^{-1}$Mpc$^{-1}$,
by detailed observations and interpretations of the anisotropies in the
CMBR~\cite{Dunkley:2008ie}, 
BBN was the only comparatively precise mean to estimate $\Omega_b$. 
As it was not clear if the "missing" dark matter was simply in form of
brown dwarfs, white dwarfs, black holes (formed from baryons), 
and/or $T\sim 10^6$K hot gas, various attempts to reconcile a BBN era 
at large $\Omega_b\sim 1$ with the observationally inferred 
light element abundances were made.
These included, for example, BBN in a baryon-inhomogeneous environment, left
over possibly due to a first-order QCD phase transition
at $T\approx 100\, $MeV, or BBN with
late-decaying particles, such as the supersymmetric 
gravitino (for reviews, cf.~\cite{Malaney:1993ah,Sarkar:1995dd,Iocco:2008va}). 
Only continuous
theoretical efforts of this sort, and their constant "failure" to account
for large $\Omega_b$, gave way to the notion that the dark matter must be
in form of "exotic", non-baryonic
material, such as a new fundamental particle 
investigated in the present book.

\section{Standard BBN - theory}
\label{SBBNtheory}

Standard BBN (SBBN) theory is well understood and described in detail
in many modern cosmology text books. The essence of SBBN is represented 
by a set of 
Boltzmann equations that may be written in the following schematic form:
\begin{eqnarray}
\label{start}
\frac{dY_i}{dt}=-H(T)T \frac{d Y_i}{ dT } = \sum (\Gamma_{ij}Y_j+ \Gamma_{ikl}  Y_k Y_l+...),
\end{eqnarray}
where $Y_i = n_i/s$ are the time $t$ (or temperature $T$) dependent ratios
between number density $n_i$ and entropy density $s$ of light 
elements $i=\,$\h1, $n$, D,\he4, etc.; the
$\Gamma_{ij...}$ are generalized rates for element 
interconversion and decay that can be 
estimated by experiments and/or theoretical calculations, 
and $H(T)$ is the temperature-dependent Hubble expansion rate.
The system of equations (\ref{start}) assumes thermal equilibrium
{\em e.g.} Maxwell-Boltzmann distributions for nuclei, which is an excellent
approximation maintained by frequent interactions with the numerous
$\gamma$'s and $e^{\pm}$'s in the plasma.
%where $Y_i$ are the time (or temperature) dependent 
%entropy-weighted abundances of light 
%elements $i=~^1{\rm H},~n,~{\rm D},~\hefm...$;
%$\Gamma_{ij...}$ are the generalized rates for element 
%interconversion and decay that can be 
%taken from experiment and/or calculated theoretically, 
%and $H(T)$ is the temperature-dependent Hubble expansion rate. 
The initial conditions for this
set of equations are well-specified: for temperatures much in excess of 
the neutron-proton 
mass difference, neutron and proton abundances are equal and related to the 
baryon to entropy ratio, $Y_{neutron} \simeq Y_{proton} \simeq \fr12 n_{baryon}/s $, 
while the abundance of all other elements is essentially zero.
%The system of equations (\ref{start}) assumes a thermal {\em e.g.} 
%Maxwell-Boltzmann distribution of nuclei over their energies, and one can show that the 
%non-thermal effects introduced by the energy release in fusion reactions is small. 
At temperatures relevant to BBN, the baryonic contribution to the
Hubble rate is minuscule, and $H(T)$ is given by the standard 
radiation-domination formula:
\bee
\label{Hubble}
H(T) = T^2\times \left(\fr{8\pi^3 g_* G_N}{90}\right)^{1/2} ,~~
{\rm where} ~ g_* = g_{boson} +\fr78g_{fermion}\, ,
\ee
with the $g$'s denoting the excited relativistic degrees of freedom. 
This expression needs to be interpolated in a 
known way across the brief epoch of the electron-positron 
annihilation, after which the photons become slightly hotter than neutrinos 
and $H(T) \simeq T_9^2/(178~{\rm s})$, where $T_9$ is the photon temperature in units of 
$10^9$K. A number of well-developed integration routine that go back to an important 
work of Wagoner, Fowler, and Hoyle~\cite{Wagoner:1966pv}, allow to solve 
the BBN system of equations 
numerically and obtain the freeze-out values of the light elements. 
A qualitative "computer-free" insight to these solutions 
can be found in {\em e.g.} Ref. \cite{Mukhanov:2003xs}. 

In a nutshell, SBBN may be described as follows.  
After all weak rates fall 
below the Hubble expansion rate, the neutron-to-proton ratio
freezes to $\sim 1/6$, 
subject to a slow further decrease to $\sim 1/7$ by $T_9\simeq 0.85$
via neutron decay and out-of-equilibrium weak conversion. 
At this point, to a good 
approximation, all neutrons available will be incorporated
into \he4, since it is the light element with the highest 
binding energy per nucleon. Synthesis of \he4, and all
other elements, has to await the presence
of significant amounts of D (the "deuterium bottleneck"). This
occurs rather late, at $T_9\simeq 0.85$, since at higher $T_9$ the fragile D 
is rapidly photodisintegrated by the multitude of CMBR photons. 
At $T_9\simle 0.85$ the fairly complete nuclear burning
of all D then results in only trace amounts {\em $O(10^{-5})$} 
of D (and \he3) being leftover after SBBN has ended. 
Elements with nucleon number $A > 4$ are even less produced
due to appreciable Coulomb barrier suppression at such low $T_9$, resulting
in only {\em $O(10^{-10})$} of \li7, and abundances of other isotopes
even lower.
SBBN terminates due to the combination of a lack of free neutrons
and the importance of Coulomb barriers at low $T$. 
In the following a few more details are given:

{\em $O(0.1)$ abundances: {\rm \hef}}.  The \he4 mass fraction $Y_p$ 
is dependent on the {\em timing} of major 
BBN events, such as the neutron-to-proton freeze-out at 
$T\simeq 0.7$ MeV, post-freezeout
neutron depletion before the deuterium bottleneck, and the 
position of this bottleneck itself
as a function of temperature. Consequently, $Y_p$ is  
dependent on such well-measured quantities as Newton's constant, the 
neutron-proton mass difference, neutron lifetime, 
deuterium binding energy, and to a much lesser degree on less
precisely known values for the nuclear reaction rates. 
This sensitivity to the timing of the BBN events makes \he4 an 
important probe of the Hubble 
expansion rate, and of all possible {\em additional} non-standard 
contributions that could modify it. 
The SBBN predicts $Y_p$ with an impressive 
precision, $Y_p = 0.2486 \pm 0.0002$, 
where we use the most recent evaluation \cite{Cyburt:2008kw}. 

{\em $O(10^{-5})$ abundances: {\rm D} and {\rm \het}}. Deuterium and \het\ 
BBN predictions are 
more sensitive both to nuclear physics and to $\eta_b$ input. 
Reactions involving 
these elements are well measured, and with the current WMAP input SBBN is 
capable of making fairly precise predictions
of these abundances: 
D/H = $2.49\pm 0.17\times 10^{-5}$;
$\hetm/{\rm H } =(1.00\pm 0.07)\times 10^{-5}$. 

{\em $O(10^{-10})$ abundances: {\rm \lisv}}. Among all observable BBN 
abundances, \lisv\ is the most sensitive to the $\eta_b$ and nuclear physics 
inputs. The actual observable that BBN predicts is the 
combined abundance of \lisv\ and \bes, as later in the course of the 
cosmological evolution \bes\ is transformed into \lisv\ via electron capture. 
At the CMBR-measured value of the baryon-to-photon ratio $\eta_b$, more than 
90\% of primordial lithium is produced in the form of \bes\ in the radiative capture process,
\hef+\het$\to$\bes+$\gamma$. As the rate for this process per each \het\ nucleus 
is much slower than the Hubble rate, the output of \bes\ is 
almost linearly dependent on the 
corresponding $S$-factor for this reaction. With recent improvement in 
its experimental determination \cite{NaraSingh:2004vj,Gyurky:2007qq,Brown:2007sj}, the current $\sim$15\% 
accuracy prediction for \bes+\lisv\ stands at 
$5.24^{+0.71}_{-0.67}\times 10^{-10}$~\cite{Cyburt:2008kw}. 

{\em $O(10^{-14})$ and less abundances: {\rm \lisx} and $A\ge 9$ elements.}
\lisx\ is formed in the BBN reaction
\bee
\label{sbbnLi6}
\hefm + {\rm D} \to \lisxm +\gamma,~~~~~Q= 1.47 {\rm MeV}
\ee
which at BBN temperatures is $\sim$ four orders of magnitude suppressed relative to other 
radiative capture reactions such as \hef+$^3$H$\to\lisvm+\gamma$, and $\sim$ seven/eight orders 
of magnitude suppressed relative to other photonless nuclear rates. The reason for the extra 
suppression is in a way accidental: it
comes from the same charge to mass ratio for \hef\ and D, which inhibits the E1 transition,
making this radiative capture extremely inefficient. This results in $O(10^{-14})$ 
level prediction for primordial \lisx\, which is well below the detection capabilities.
Heavier elements with $A\ge 9$ such as \ben, $^{10}$B and $^{11}$B are never made in any significant 
quantities in the SBBN framework, and the main reason for 
that is the absence of stable $A=8$ nuclei, 
as \beet\ is underbound by 92 keV and decays to two $\alpha$.  
%Thus, \lisx\ and heavier elements with $A\ge 9$ are never produced in BBN in observationally 
%significant quantities, and detection of such elements in significant quantitites 
%may signal the presence of new physics. 

\section{Observed light element abundances}
\label{observations}

In the following, we will briefly discuss the observationally inferred
light elements, element by element. Here the discussion will also include
isotopes, or isotope ratios, such as \li6, \be9, and \he3/\h2, 
which are not always considered
in SBBN, but which are very useful to constrain deviations from SBBN.

\subsection{\he4}

The primordial \he4/H ratio is inferred from observations of hydrogen- 
and helium- emission lines in extragalactic low-metallicity 
HII-regions and compact blue galaxies, illuminated by young star clusters. 
Two particular groups have performed
such analysis for years now, with their most recent results
$Y_p\approx 0.2477\pm 0.0029$~\cite{Peimbert:2007vm} and 
$Y_p\approx 0.2516\pm 0.0011$~\cite{Izotov:2007ed}. 
These estimates are (surprisingly) considerably
larger than earlier estimates by both groups
(i.e. $0.239$ and $0.242$, respectively), explained in large parts
by a new estimate for HeI emissivities~\cite{Porter:2006gd}. 
Other differences with
respect to older studies, and/or between the two new studies themselves, 
are the adopted rates for collisional excitation of H- (He-) emission lines,
corrections for a temperature structure in these galaxies 
("temperature variations"), corrections for the presence of
neutral \he4 ("icf - ionisation corrections"), as well as
corrections for throughs in
the stellar spectra at the position of the \he4- (H-) emission lines
("underlying stellar absorption"). All of these may have impact on the
$\simge 1\%$ level. This, as well as the comparatively large change from
earlier estimates (coincidentally going into the direction of agreement
with the SBBN prediction of $Y_p\approx 0.248$), implies that a conservative
estimate $Y_p\approx 0.249\pm 0.009$~\cite{Olive:2004kq} 
(see also $Y_p\approx 0.250\pm 0.004$)~\cite{Fukugita:2006} of the error bar, 
maybe more appropriate when constraining perturbations of SBBN.

\subsection{D}

For the observational determination of primordial D/H-ratios
high-resolution observations of low-metallicity quasar absorption line
systems (QALS) are employed 
(cf.~\cite{Burles:1997ez,Levshakov:2001xi,Crighton:2004aj,O'Meara:2006mj,Pettini:2008mq}).
QALS are clouds of partially neutral gas which fall on the line of sight 
between the observer and a high-redshift quasar. The
neutral component in these clouds
yields absorption features, for example, at the redshifted position of 
the Lyman-$\alpha$ wavelength. For the very rare QALS of sufficiently simple
velocity structure, one may compare the absorption at the Lyman-$\alpha$
position of H with that of D (shifted by 81 km s$^{-1}$) to infer a
D/H ratio. Here the low metallicity of these QALS is conducive to 
make one believe
that stellar D destruction in such clouds is negligible. Currently there
exist only about $6-8$ QALS with D/H determinations. When averaged
they yield typical 
$2.68-2.82\times 10^{-5}$~\cite{O'Meara:2006mj,Fields:2006ga,Steigman:2007xt,Pettini:2008mq} 
for the central value, with inferred statistical $1\sigma$ error bars
of $0.2-0.3\times 10^{-5}$, comparing favorably to the SBBN prediction
of $2.49\pm 0.17\times 10^{-5}$~\cite{Cyburt:2008kw} at the WMAP inferred 
$\eta_b$. Nevertheless, as an important cautionary remark,
the various inferred D/H ratios in QALS show a spread considerably
larger than that expected from the above quoted error bars only. 
This is usually a sign of 
the existence of unknown systematic errors. 
Until these systematics are better understood
should primordial values as high as 
D/H$\approx 4\times 10^{-5}$ therefore not considered to be ruled out.

\subsection{\he3/D}

Observational determinations of \he3/H-ratios are possible
within our galaxy which is chemically evolved. 
The chemical evolution of \he3 is, however, rather
involved, with \he3 known to be produced in some stars and destroyed in 
others. Furthermore, any D entering stars will be converted
to \he3 by proton burning. The net effect of all this is an observed
approximate constancy of 
(D+\he3)/H$\,\approx 3.6\pm 0.5\times 10^{-5}$~\cite{Geiss:2007} over the last 
few billion years in our galaxy. Whereas the relation of galactic observed
\he3/H ratios to the primordial one is obscure, the ratio of \he3/D 
as observed in the presolar nebulae is invaluable in 
constraining perturbations of SBBN. 
This ratio $0.83^{+0.53}_{-0.25}$~\cite{Geiss:2007} 
(where the error bars are obtained when using the independent $2\sigma$
ranges of \he3/H and D/H)
provides a firm upper limit on the 
primordial \he3/D~\cite{Sigl:1995kk}. This is because 
\he3 may 
be either produced or destroyed in stars, while D is always destroyed, 
such that the cosmological \he3/D ratio may only grow in time.

\subsection{\li7}

\li7/H ratios may be inferred from observations of absorption lines
(such as the 6708A doublet) in the atmospheres of low-metallicity
galactic halo stars. When this is done for stars at low metallicity $[Z]$,
\li7/H ratios show a well-known anomaly (with respect to other elements),
i.e. \li7/H ratios are constant over a wide range of (low) [Z] and 
some range of temperature (the "Spite plateau"). 
As most elements are produced by stars 
and/or cosmic rays, which themselves produce metallicity, the 
\li7 Spite plateau is believed to 
be an indication of a primordial origin of this isotope. 
This interpretation
is strengthened by the absence of any observed scatter in the \li7 abundance
for such stars. There have been several observational determinations of
the \li7 abundance on the Spite plateau. Most of them fall in the range
\li7/H$\,\approx 1-2\times 10^{-10}$ such as 
$1.23^{+0.68}_{-0.32}\times 10^{-10}$~\cite{Ryan:1999jq,Hosford:2008rs}
and $1.1-1.5\times 10^{-10}$~\cite{Asplund:2005yt}, with some being somewhat
higher such as $2.19\pm 0.28\times 10^{-10}$~\cite{Bonifacio:2002yx}. 
Here differences may be
due to differing methods of atmospheric temperature estimation. These values
should be compared to the SBBN prediction
$5.24^{+0.71}_{-0.67}\times 10^{-10}$~\cite{Cyburt:2008kw} 
(with $1\sigma$ error estimates),
clearly indicating a conflict which is often referred 
to as the "lithium problem". It is essentially ruled out that 
this problem be solved by, either, an erroneous atmospheric temperature
determination, or significant changes in \li7 producing/destroying nuclear
SBBN rates. There remain only two viable possibilities of a resolution to 
this statistically significant $(4-5)\sigma$ problem. First, it is conceivable
that atmospheric \li7 has been partially destroyed in such stars 
due to nuclear burning in the stellar interior. 
Though far from understood, one may indeed construct (currently ad hoc)
models which deplete \li7 by a factor $\approx 2$ in
such stars, while respecting all 
other observations~\cite{Richard:2004pj,Korn:2006tv}. Second, it is possible that the
lithium problem points directly towards physics beyond the 
SBBN model, possibly connected to the production of the dark matter
(cf. Section~\ref{decay}).

\subsection{\li6 and \be9}

The isotope of \li6 is usually not associated with BBN, as its
standard BBN production \li6/H$\sim 10^{-14}$ is very low. However, 
the smallest deviations from SBBN may already
lead to important cosmological \li6
abundances. It is therefore interesting that the existence of \li6
has been claimed in about $\sim 10$ low-metallicity 
stars~\cite{Asplund:2005yt}, with,
nevertheless, each of these observations only at 
the $2-4\sigma$ statistical significance level. 
Asplund {\it et al.} infer an average
of \li6/\li7$\approx 0.044$ 
(corresponding to \li6/H$\approx 6\times 10^{-12}$) for their star sample,
whereas Cayrel {\it et al.} infer \li6/\li7$\approx 0.052\pm 0.019.$ for the 
star HD84937~\cite{Cayrel:1999xx}. 
Such claims, if true,
would be of great interest, as the inferred \li6/\li7 in very low
metallicity stars is exceedingly hard to explain by cosmic ray 
production~\cite{Prantzos:2005mh}, 
though in situ production in stellar flares may be 
conceivable~\cite{Tatischeff:2006tw}. 
Moreover, the \li6 observations seem to be consistent with a
plateau structure at low metallicity as expected
when originating right from BBN. However, recent work~\cite{Cayrel:2007te} 
has cast significant shadow
over the claim of elevated \li6/\li7 ratios at low [Z]. 
Similar to \li7,
\li6 is inferred from observations of 
atmospheric stellar absorption features. Unlike in the case of D and H
in QALS, the absorption lines of \li7 and \li6 are always
blended together. \li6/\li7 ratios may therefore be obtained only by 
observations of a
minute asymmetry in the 6708 line. Such asymmetries could be
due to \li6, but may also be due to asymmetric convective motions in the
stellar atmospheres. The analysis in Cayrel {\it et al.}~\cite{Cayrel:2007te}
prefers the latter explanation.

Unlike the case of \lisx, detection of \ben\ in many stars at low 
metallicities is not 
controversial. Observations of \ben\ \cite{Primas:2000gc,Boesgaard:2005pf} are far 
above the $O(10^{-18})$ SBBN prediction, 
and exhibit linear correlation with oxygen, clearly indicating its secondary (spallation)
origin \cite{Fields:2004ug}. The lowest level of detected \ben/H is at 
$\sim\,$few$\times 10^{-14}$, which translates into the 
limit on the primordial fraction at $2\times 10^{-13}$ \cite{Pospelov:2008ta}, 
assuming no significant depletion of \ben\
in stellar atmospheres.

\section{Cascade nucleosynthesis from energy injection}
\label{cascade} 

%\begin{figure}
%\begin{center}
%\includegraphics[bb=101 142 504 652, width=.7\textwidth,
%clip=true, keepaspectratio]{chart-2.ps}
%\end{center}
%\caption{Diagram illustrating the effects of relic particle decay on the
%plasma, from Ref.~\cite{Kawasaki:2004qu}.}
%\label{fig:chart}
%\end{figure}

\begin{figure}
\begin{center}
\includegraphics[width=.65\textwidth]{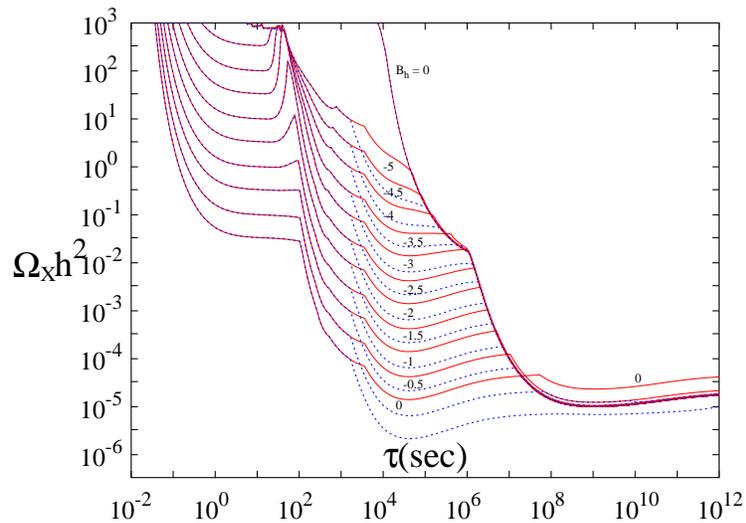}
\end{center}
\caption{Constraints on the abundance 
$\Omega_Xh^2$ of relic
particles decaying at $\tau_X$ assuming $M_X = 100\,$GeV for the 
particle mass. The most stringent limits are given, 
from early to late times, by
\he4, D, \li6, and \he3/D overproduction, respectively. The various 
lines are for different ${\rm log_{10}}B_h$, as labeled, 
where $B_h$ is the hadronic branching ratio. 
>From Ref.~\cite{Jedamzik:2006xz}.
}
\label{fig:constraint}
\end{figure} 

The possibility that BBN may be significantly perturbed by the presence
of energetic, non-thermal SM particles in the plasma 
has first received detailed attention
in the 1980s~\cite{Ellis:1984er,Levitan:1988au,Dimopoulos:1987fz,Reno:1987qw,
Dimopoulos:1988ue,Ellis:1990nb,Khlopov:1993ye,Kawasaki:1994sc}. 
Though much of the pioneering work
had been done, only recently the first fully realistic 
calculations of coupled thermal 
nuclear reactions and non-energetic phenomena 
have been presented~\cite{Jedamzik:2004er,Kawasaki:2004yh,Kawasaki:2004qu,Cyburt:2006uv}. 
Energetic particles may be injected as products of the decay or annihilation
of relic non-SM particles, or via perhaps more exotic mechanisms such as evaporation of
primordial black holes or supersymmetric Q-balls. The injected energetic 
photons $\gamma$'s, electron/positrons $e^{\pm}$'s, neutrinos $\nu$'s 
muons $\mu^{\pm}$'s, pions $\pi$'s, nucleons and antinucleons 
$N$'s and $\bar{N}$'s, gauge bosons $Z$'s and $W^{\pm}$'s, etc. 
may be considered
as the "cosmic rays" of the early Universe. 
In contrast to their present day
counterparts, and with the exception of neutrinos,
these early cosmic rays thermalize rapidly within a small
fraction of the Hubble time $H^{-1}(T)$ for 
all cosmic temperatures above $T\sim 1\,$eV.
This, of course, happens only after all unstable species (i.e. $\pi$'s, $\mu$'s
$Z$'s and $W^{\pm}$'s) have decayed leaving only $\gamma$'s, $e^{\pm}$'s, 
$\nu$'s and $N$'s. Many of the changes in BBN light-element production occur
during the course of this thermalization. One often distinguishes between
hadronically ($\pi$'s, $N$'s, and $\bar{N}$'s) and
electromagnetically ($\gamma$'s, $e^{\pm}$'s) interacting particles,
mainly because the former may change BBN yields at times as early as 
$\tau\simge 0.1\,$sec (i.e. $T\simle 3\,$MeV), whereas the latter only
have impact for $\tau\simge 10^5\,$sec (i.e. $T\simle 3\,$keV).
In the following we summarize the most important interactions and outline the
impact of such particles on BBN.  For hadronically interacting particles these effects include:
\begin{enumerate}

\item 
$\pi^{\pm}$'s may cause charge exchange, i.e. $\pi^- + p\to \pi^0 + n$ 
between $1\,{\rm MeV}\simge T\simge$ $300\,{\rm keV}$ thereby creating extra
neutrons after $n/p$ freezeout and increasing the helium mass fraction $Y_p$.

\item
Antinucleons $\bar{N}$ injected in the primordial plasma 
preferentially annihilate on protons, thereby raising the
effective $n/p$-ratio and increasing $Y_p$.

\item 
At higher temperatures, neutrons $n$'s completely thermalize through magnetic
moment scattering on $e^{\pm}$ ($T\simge 80\,$keV), whereas protons $p$'s
do so through Coulomb interactions with $e^{\pm}$ and Thomson 
scattering off CMBR photons ($T\simge 20\,$keV). 
%In neither case are abundance yields changed.  %% I did not understand this sentence - MP. 
Any extra neutrons at $T\sim 40$ keV may lead to an 
important depletion of \bes. 

\item
At lower temperatures, both, energetic neutrons and protons may spall \he4,
e.g. $n +{\rm {}^4He\to {}^3H} + p + n +(\pi$'s), or
$n +{\rm {}^4He\to {}D} + p + 2n +(\pi$'s). Both reactions are important as
they may either increase the \h2 abundance or lead to \li6 formation 
via the secondary  non-thermal reactions of energetic ${}^3$H(\het) on ambient $\alpha$'s. 
\end{enumerate}
The main features of electromagnetic injection are:
\begin{enumerate}
\item 
Energetic $\gamma$'s may pair-produce on CMBR photons, i.e. 
$\gamma +\gamma_{\rm CMBR}\to e^- + e^+$
as long as their energy is above the threshold $E_C\approx m_e^2/22T$ for
this process. The created energetic $e^{\pm}$ in turn inverse Compton scatter,
i.e. $e^{\pm}+\gamma_{\rm CMBR}\to e^{\pm}+\gamma$, to produce 
further $\gamma$'s. Interactions with CMBR photons completely dominate 
interactions with matter due to the exceedingly small cosmic baryon-to-photon
ratio $\eta$.

\item 
Only when $E_{\gamma}\simle E_C$ do interactions with matter
become important. These include Bethe-Heitler pair production
$\gamma + p({\rm {}^4He})\to p({\rm {}^4He})+e^++e^-$ and Compton scattering
$\gamma + e^-\to \gamma + e^-$ off plasma electrons, as well as
photodisintegration (see below).

\item
A small fraction of $\gamma$'s with $E_{\gamma}\simle E_C$ may
photodisintegrate first D at $T\simle 3\,$keV, when $E_C$ becomes
larger that $E_b^{\rm D}\approx 2.2\,$MeV, the D binding energy, 
and later \he4, at $T\simle 0.3\,$keV since 
$E_b^{\rm 4He}\approx 19.8\,$MeV. Such processes may cause
first, D destruction, and later, D and \he3 production, and more 
importantly \he3/D overproduction. They may also lead to \li6 production. 
\end{enumerate} 
%These interactions are also displayed in the chart of figure~\ref{fig:chart}.

In the context of non-thermal energy injection related to 
particle dark matter, there are two very important processes
that have profound impact on \lisx\ and \lisv\ abundances and deserve further comments. 
Energetic \het\ and ${}^3$H produced via electromagnetic or hadronic energy 
injection ({\em i.e.} via spallation or photodisintegration) 
provide the possibility of efficient production of 
\li6 via the non-thermal nuclear reactions on thermal \he4:
\bee
\label{nonthermal}\!\!\!\!\!\!\!\!\!\!\!\!\!
^3{\rm H} + \hefm \to \lisxm + n, ~~  Q = -4.78{\rm MeV};~~
\hetm + \hefm \to\lisxm +p, ~~ Q = -4.02{\rm MeV}.
\ee 
For energies of projectiles $\sim$10 MeV, the cross sections 
for these nonthermal processes are on the order of 100 mbn,
and indeed $10^{7}$ times larger than the SBBN cross section 
for producing \lisx\ . 
This enhancement figure underlines the \lisx\ sensitivity to non-thermal 
BBN, and makes it an important probe of energy injection
mechanisms in the early Universe.

Another important aspect of the nonthermal BBN is the possibility to 
alleviate the tension between the Spite plateau value and the predicted abundance of 
\lisv, {\em e.g.} "solve the \lisv\ problem". 
To achieve that the energy injection should
occur in the temperature  interval 
$60\, {\rm keV}\simge T\simge 30\, {\rm keV}$, {\em i.e.} 
during or just after \bes\ synthesis . 
The essence of this mechanism consists in the injection of $\simge 10^{-5}$ 
neutrons per baryon that will enhance \bes$\to$\lisv\ interconversion followed by the 
$p$-destruction of \lisv\ via the thermal reaction 
sequence~\cite{Jedamzik:2004er}: 
\begin{equation}
n+{^7\rm Be}\to p +{^7\rm Li};~~
\,\, p+{^7\rm Li}\to {^4\rm He} + {^4\rm He}\, .
\label{Li}
\end{equation}
Note that this is the same mechanism that depletes \bes\ in SBBN, 
but with elevated neutron concentration due to the hadronic energy injection. 
This mechanism of depleting \bes\ is tightly constrained by the deuterium abundance, 
as extra neutrons could easily overproduce D. 

Fig.~\ref{fig:constraint} summarizes constraints on abundance 
vs lifetime of relic
decaying particles. It is convenient to measure the abundance in terms of $\Omega_Xh^2$,
the present day fraction of total energy density if these particles were to remain stable. 
This quantity relates to the in the literature frequently used
$\zeta = n_X M_X/s$ via $\zeta = 3.6639\times 10^{-9}{\rm GeV}\Omega_Xh^2$
where $n_X$ is particle number density, $M_X$ is particle mass, 
and $s$ is entropy.
It is seen that constraints get increasingly more
stringent when the lifetime $\tau_X$ increases, implying also, that under generic 
circumstances, the production of dark matter $X$ (with $\Omega_Xh^2\sim 0.1$)
by the decay of a parent particle $Y\to X +...$ at $\tau_X\gg 10^3$sec is 
extremely problematic, if at all possible. 

\section{Residual dark matter annihilation during BBN} 
\label{annihilation}

\begin{figure}
\begin{center}
\includegraphics[width=.65\textwidth]{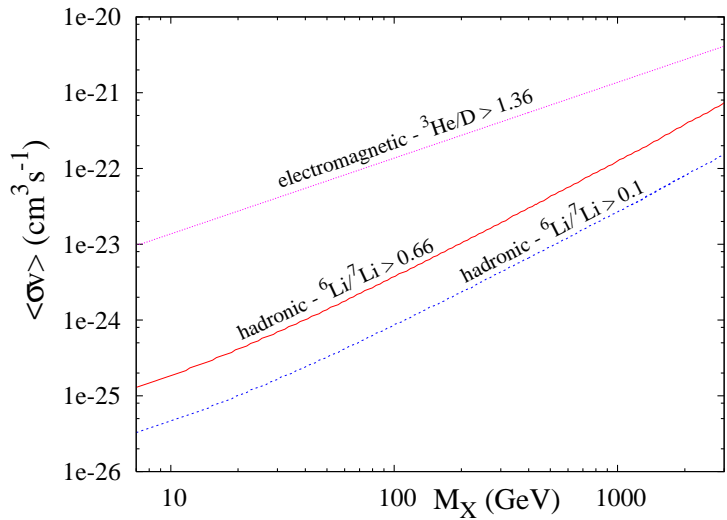}
\end{center}
\caption{Upper bound on the annihilation cross section
of particle dark matter of mass $M_{\chi}$ from BBN. Here the upper 
line assumes annihilation into only electromagnetically interacting
particles, whereas the two lower lines assume annihilation into
a light quark-anti-quark pair. Adopted limits on the light element abundances
are as indicated in the figure. }
\label{fig:ann1}
\end{figure}

\begin{figure}
\begin{center}
\includegraphics[width=.65\textwidth]{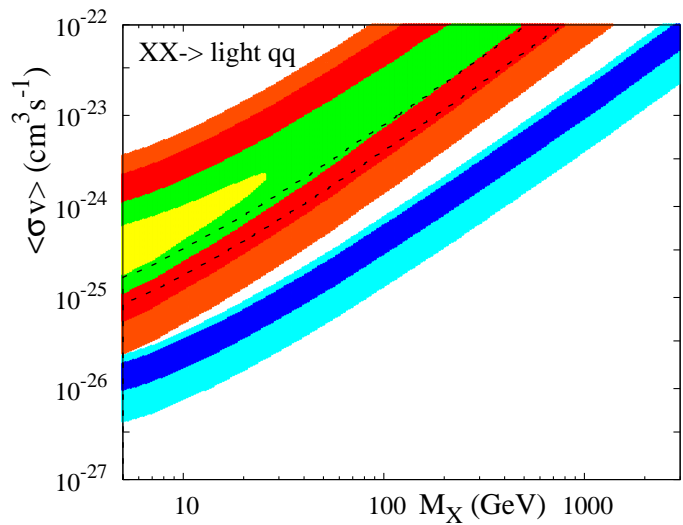}
\end{center}
\caption{Dark matter annihilation rate versus dark matter mass.
The blue band shows parameters where \li6 due to
residual dark matter annihilation
may account for the \li6 abundance as inferred in HD84937
(\li6/\li7$\approx 0.014-0.09$ at 2-$\sigma$), whereas the 
orange-red-green-yellow region shows where \li7 is efficiently destroyed
i.e. \li7/H$<1.5,2,3,$
and $4\times 10^{-10}$, respectively. Above the lower (upper) dashed
line D/H exceeds $4\times 10^{-5}$ ($5.3\times 10^{-5}$), such that
parameter space above the upper dashed line is ruled out by 
D overproduction. Scenarios between this line and the upper edge of the 
blue band are problematic since severely overproducing \li6.
Dark matter annihilation into light quarks has been assumed.
}
\label{fig:ann}
\end{figure} 

Many dark matter candidates $X$ may be ultimately visible in our Galaxy
due to the cosmic rays they inject induced by 
residual $XX$ self-annihilations. This is, for example, the case for 
supersymmetric neutralinos, provided that their 
annihilation products can be distinguished from 
astrophysical backgrounds. 
Residual annihilation events in the early Universe may also be of importance 
as they may 
lead to cosmologically significant \li6 abundances~\cite{Jedamzik:2004ip} 
induced by the non-thermal nuclear reaction discussed in 
Section~\ref{cascade}. 
Given an $X$ annihilation rate $\langle\sigma v\rangle$  and
$X$-density $n_X$ one may
determine the approximate fraction $f_X$ of $X$ particles which annihilate in the
early Universe at temperature $T$
\begin{equation}
f_X \approx \frac{1}{n_X}\frac{{\rm d}n_X}{{\rm d} t}\Delta t_H
    \approx \langle\sigma v\rangle \frac{n_X}{s}\,\, sH^{-1}
%\Delta t_H
 %   \approx \langle\sigma v\rangle_{0}
  %  \frac{n_X}{s}\,\, sH^{-1}\biggl(\frac{v}{v_{0}}\biggr)^n
\label{eq:fX}    
\end{equation}
where $\Delta t_H\approx H^{-1}  = (90 M_{pl}^2/\pi^2 g T^4)^{1/2}$ 
is the
characteristic Hubble time at $T$, $g$ is the appropriate particle
statistical weight, $s = 4\pi^2/90\, g T^3$ is radiation entropy,
and $\langle ... \rangle$ denotes a thermal average,
which can be taken once the velocity dependence of $\sigma v$ is 
specified. Many scenarios for the production of dark matter envision a stable
self-annihilating particle, typically a weakly-interacting massive particle or WIMP,  
whose final asymptotic abundance is given by
its annihilation rate. Straightforward considerations of thermal WIMP freeze-out 
require the annihilation rate at $T_f^{th} \simeq 0.05 m_X$ to be 
$\langle\sigma v\rangle_{f}^{th}\approx 
1{\rm pbn} \times c = 3\times 10^{-26}{\rm cm^3s^{-1}}$ 
if the $X$-particle is to be the dominant component of 
dark matter, $\Omega_Xh^2\approx 0.1$. 
%We denote this benchmark 
%size of cross section by 
%$(\sigma v)_0$. 
Less straightforward but still plausible scenarios that include the 
non-thermal production of dark matter, 
{\em e.g.} via evaporation of Q-balls and/or decay of relic particles $Y\to X + ...$
with subsequent $X$ self-annihilation, may require
$\langle\sigma v\rangle_f$ well in excess 
of $(\sigma v)_{f}^{th}$ for $\Omega_Xh^2\approx 0.1$.
We concentrate on the WIMP example, and 
parametrize the velocity dependence of $\langle\sigma v\rangle$ by 
$\langle\sigma v\rangle = (\sigma v)_0 S(v)$,
with the chosen normalization $(\sigma v)_0 =\langle\sigma v\rangle_f$.
%In both of these cases
%Eq.~\ref{eq:fX} may be used with $f_X\approx 1$ at freeze-out "$f$" of the
%final abundance. The same equation, however also applies at lower $T$,
%when only residual annihilations $f_X\ll 1$ take place. 
%Dividing Eq.~\ref{eq:fX} at low $T$ by Eq.~\ref{eq:fX} evaluated
%at freeze-out one thus finds
We are interested in finding the fraction of 
annihilating WIMP particles at $ T < 10\,$keV, 
a temperature scale below which \lisx\ is no longer susceptible to 
nuclear burning (destruction).
Exploiting (\ref{eq:fX}) at the freeze-out 
temperature $T_f$, where $f_X(T_f)\approx 1$, 
as well as at an arbitrary other $T\ll T_f$, we obtain
\begin{equation}
f_X(T) \approx \biggl(\frac{g(T)}{g(T_{f})}\biggr)^{1/2}
      \biggl(\frac{T}{T_{f}}\biggr)
      \frac{\langle S(v) \rangle_T}{\langle S(v) \rangle_{T_f}}       
%\biggl(\frac{g(T)}{g(T_{f})}\biggr)^{1/2}\biggl(\frac{T}{T_{f}}\biggr)^{1+\fr{n}{2}}
\label{fX(T)}      
\end{equation}
for $f_X\ll 1$. Several generic options are possible for the temperature scaling of the 
the $S$-ratio in (\ref{fX(T)}).  If the $s$-wave annihilation is mediated by 
short-distance physics 
and occurs away from sharp narrow resonances, 
$\langle S(v) \rangle_T=\langle S(v) \rangle_{T_f} =1$, where the
second equality is due to our chosen normalization.
Using  this conservative assumption and Eq.~(\ref{fX(T)}), for a WIMP of mass
$M_X = 100\,$GeV, so that $g(T)/g(T_f) \simeq 0.1$,   one finds
that only a small fraction, $f_X\approx 6\times 10^{-7}$, of $X$-particles 
has a chance to
annihilate at $T \simeq 10\,$keV and below. 
Nevertheless, even this tiny fraction is still sufficient to 
produce a \li6 abundance of \li6/H$\approx 1.6\times 10^{-12}$.

Existence of attractive Coulomb-like force of some strength $\alpha'$ in the 
WIMP sector may lead to a significant enhancement 
of annihilation at low temperatures/velocities\cite{Hisano:2003ec}, possibly
leading to a much higher yield of \lisx.  
In this case the Sommerfeld-like 
scaling $\sigma v \sim (\pi \alpha'/v)[1-\exp(-\pi \alpha'/v)]^{-1}$, 
enhances the  annihilation at small $v\simle \pi\alpha'$, 
i.e. $\langle S(v) \rangle_{T} \simeq \pi\alpha'/v$.
This leads to a $\sim T^{-1/2}$ scaling 
of $\langle S(v) \rangle_{T}$ in (\ref{fX(T)}) when X-particles are still
in thermal equilibrium with the plasma. 
After they have dropped out of thermal
equilibrium, 
%(often at $T\simle 100-1\,$MeV), 
$\langle S(v) \rangle_{T}$ falls even more rapidly
as $\sim T^{-1}$, with the net effect that weak mass scale X-particles 
usually have
much smaller velocities at the end of BBN than in the Milky Way. 
Similarly, the presence of narrow 
resonances just above the 
$XX$ annihilation threshold may drastically boost 
the annihilation at low energies. 
Both mechanisms of enhancing the annihilation have been 
widely discussed (see {\em e.g.} \cite{Pospelov:2008jd}) in an attempt to link some  
cosmic-ray anomalies to dark matter annihilation, 
as for example an elevated positron fraction 
$e^+/(e^-+e^+)$  observed by PAMELA instrument \cite{Adriani:2008zr}.
 
Keeping the annihilation rate as a free parameter, Fig.~\ref{fig:ann1}
shows the upper limit on the effective annihilation cross section imposed
by BBN. Here electomagnetically- (upper line) and hadronically- 
(lower lines) annihilating particle DM has been considered. The former is
mostly constrained by overproduction of \he3/D at $T\approx 0.1\,$keV,
while the latter by \li6/\li7 overproduction at $T\approx 10\,$keV,
such that the effective annihilation cross section refer to
$\langle\sigma v\rangle$ at those temperatures.
Due to the possibility of \li6 destruction a fairly 
conservative \li6/\li7$<0.66$
constraint has also been considered. It is seen that much \li6 may be
produced by hadronic annihilations.
Fig.~\ref{fig:ann} shows  
dark matter parameters which lead to the production
of a \li6/\li7 ratio as claimed to be observed in the star HD84937
at $1\sigma$ and $2\sigma$ (dark blue and light blue), respectively. 
Here a completely hadronic 
$XX\to q\bar{q}$ annihilation has been employed, 
an assumption 
which could be further confronted with the constraints on 
antiproton fluxes in our galaxy. 
Electromagnetic annihilations, as often implied in recent 
dark-matter interpretations 
of PAMELA\cite{Adriani:2008zr}, may also lead to \li6, but only annihilations 
below $T\simle 0.3\,$keV may do so efficiently.
Note, that the figure already implicitly assumes a
factor $\sim 3-4$ stellar destruction of \li7 (and \li6) to solve the
lithium problem.
It is therefore found that weak-scale mass dark matter
particles, if fairly light, and if annihilating into hadronically 
interacting particles, may account for all of the 
observed \li6 in HD84937. The figure also shows by the 
orange-red-green-yellow areas the dark matter parameters which would lead to
a significant \li7 destruction due to residual dark matter annihilations,
with the nuclear destruction mechanism described in the previous section.
Since those regions are much above the \li6 band, the possibility of a factor of 2 or more 
depletion in \li7 is severely constrained by \li6 overproduction. 
We note in passing that our constraints are significantly more conservative than those found in 
Ref.~\cite{Hisano:2008ti}.
It is intriguing to realize that primordial \li6 production 
by residual (hadronic) dark matter annihilations 
dominates standard BBN \li6 production
for annihilation rates as small as
$\langle\sigma v\rangle\sim 10^{-27}{\rm cm^3s^{-1}}$, well 
below $\langle\sigma v\rangle_{f}^{th}$. It is thus possible
that in the most primeval gas clouds and in the oldest stars the
bulk of \li6 is due to dark matter annihilations. Unfortunately, \li6 
abundances as low as \li6/H $\simle 10^{-12}$ are difficult to observe.

\section{Catalyzed BBN (CBBN)} 
\label{catalysis} 

The idea of particle physics catalysis of nuclear reactions 
goes back to the 1950s, and muon-catalyzed
fusion has been a subject of active theoretical and 
experimental research in nuclear physics.
In recent years 
there has been a significant interest towards a possibility of
nuclear catalysis by hypothetical negatively charged particles 
that live long enough to participate in nuclear reactions 
at the BBN 
time~\cite{Pospelov:2006sc,Kohri:2006cn,Kaplinghat:2006qr,Cyburt:2006uv,
Hamaguchi:2007mp,Bird:2007ge,Jittoh:2007fr,Jedamzik:2007cp,Jedamzik:2007qk,
Pospelov:2007js,Kusakabe:2007fv,Pospelov:2008ta,Kamimura:2008fx} (see also 
Ref.~\cite{DeRujula:1989fe,Dimopoulos:1989hk,Rafelski:1989pz} 
for earlier work on the subject). 
An essence of the idea is very simple:
a negatively charged massive particle that we call \xm\ gets into 
a bound state with the nucleus of mass $m_N$ and charge $Z$, forming a 
large compound nucleus with the charge $Z-1$, mass $M_X+m_N$, 
and the binding energy in the $O(0.1-1)$ MeV range. 
Once the bound state is formed, the Coulomb barrier is reduced 
signalling a higher "reactivity" of the compound nucleus with 
other nuclei. 
But what proves to be the most important effects of catalysis,
are new reaction channels which may open up and 
avoid SBBN-suppressed production
mechanisms \cite{Pospelov:2006sc}, {\em e.g.} Eq. (\ref{sbbnLi6}), thus 
clearing path to synthesis of 
elements such as \lisx\ and \ben. 
Although in this chapter we discuss the catalysis by negatively charged 
heavy relics, this is not the only option for CBBN, as for example,
strongly interacting relics may also participate and catalyze certain nuclear
reactions. 

Although the connection between dark matter and CBBN is not immediate -
after all the dark matter may not be charged - it is possible that 
dark matter particles do have a relatively long-lived
charged counterpart. One example of this kind is supersymmetry with
the lightest supersymmetric particle (LSP) the 
gravitino and the next-to-LSP (NLSP) a charged slepton 
to be examined in the 
next section. In that case the decay of the 
NLSP is tremendously delayed by the smallness of the gravitino-lepton-slepton
coupling $\sim G_N^{1/2}$. Another example in the same vein is the nearly degenerate
stau-neutralino system, in which case the longevity of the
charged stau against the decay to the 
dark matter neutralino is ensured 
as long as the mass splitting of the stau-neutralino system is below 100 MeV. 
Both, the gravitino and neutralino in these two examples represent viable 
dark matter candidates. 
A very important aspect of CBBN is that the abundance of charged particles
before they start decaying is given by their annihilation rate
at freeze-out. In most of the models their abundance is easily calculable, and 
if no special mechanisms are introduced to boost the annihilation rate, 
the abundance of charged particles per nucleon is not small, and in the 
typical ballpark of $Y_X \sim (0.001-0.1)\times m_X/{\rm TeV}$. 

{\em Properties of the bound states.}

For light nuclei participating in BBN, 
we can assume that the reduced mass of the nucleus-\xm system 
is well approximated by the nuclear mass, so that the binding energy is 
given by  $Z^2\alpha^2 m_A /2$ when the Bohr orbit is larger than the 
nuclear radius. It turns out that this is a poor approximation for 
all nuclei heavier than $A=4$, and the effect of the finite nuclear charge radius 
has to be taken into account. In Table \ref{table1} we give the 
binding energies, as well as the recombination 
temperature, defined as the temperature at which the 
photodissociation rate of 
bound states becomes smaller than the Hubble expansion rate. 
Below these temperatures bound states are practically stable,
and the most important benchmark temperatures for the CBBN are
the $T \sim 30,~8,~0.5$ keV, when (\bes\xm), (\hef\xm), and ($p$\xm)
can be formed without efficient 
suppression by the photodissociation processes.
It is important to emphasize that these properties of the bound states are 
generic to any CBBN realization: {\em i.e.} they are completely 
determined by the charge of \xm\ and electromagnetic properties of nuclei, and thus are
applicable to SUSY or non-SUSY models alike. It is also important 
to note that the 
(\beet\xm) compound nucleus is stable, which may open the path to synthesis of
$A>8$ elements in CBBN. 

\begin{table}
\begin{center}
\begin{tabular}{|c|c|c|c|c|c|c|c|}
\hline
bound state &            $a_0$[fm] &     $|E_b|$[keV] &~$T_0$[keV]~ \\ \hline\hline
p$X^-$&                 29  &                 25    & 0.6  \\ 
\hef$X^-$&            3.63 &        346     &  8.2 \\ 
%\het$X^-$&          299  &   4.81 &    1.76    &       276         &    2.50   &         267     &  6.3 \\ \hline
%\lisx$X^-$&        1343  &   1.61 &    2.22    &       930         &    3.29   &         780     &  19  \\ \hline
%\lisv$X^-$&        1566  &   1.38 &    2.33    &       990         &    3.09   &         870     &  21  \\ \hline
\bes$X^-$&            1.03 &          1350    &  32  \\ 
\beet$X^-$&           0.91 &           1430    &  34  \\ \hline
%\hef $X^{--} $ &   1589  &   1.81 &    1.94    &       1200        &      2.16 &         1150    &  28  \\ \hline\hline
\end{tabular}
\end{center}
\caption{Properties of the bound states: Bohr radius $a_0 = 1/(Z\alpha m_N)$, 
binding energies $E_b$ calculated for realistic charge radii, and 
``photo-dissociation decoupling" temperatures
$ T_0$.
}
%\vspace{-0.5cm}
\label{table1}
\end{table}

{\em Catalysis at 30 keV: suppression of} \bes.

When the Universe cools to temperatures of 30 keV, 
the abundances of deuterium, \het, \hef, \bes\ and \lisv\ are already 
close to their freeze-out values, although several nuclear processes 
remain faster than the Hubble rate. At these 
temperatures, a negatively charged relic can get into bound states with
\bes\ and form a (\bes\xm) composite object. Once this object is formed, 
some new destruction mechanisms for \bes\ appear. For models with weak currents 
connecting nearly mass-degenerate \xm-\xz\ states, a very fast internal conversion 
followed by the $p$-destruction of \lisv:
\bee
\bexm\to \lisvm\,+X^0;~\lisvm+p\to 2\alpha.
\label{intconversion}
\ee
When \xm\ $\to$\xz\, is energetically disallowed the 
destruction of \bes\ can be achieved via the following chain:
\bee
\label{pburning}
\bexm + p \to (^{8}{\rm B}X^-) +\gamma:~ (^{8}{\rm B}X^-) \to (^{8}{\rm Be}X^-) + e^+\nu,
\ee
which is much enhanced by the atomic resonances in 
the $\bexm $ system \cite{Bird:2007ge}. 

The rates for both mechanisms may be faster than the Hubble rate, possibly leading to 
a sizable suppression of \bes\ abundance {\em if} (\bes\xm) bound states are efficiently forming. 
In other words, (\bes\xm) serves as a bottleneck for the CBBN depletion of \bes.
The recombination rate per \bes\ nucleus leading to (\bes\xm) is given by the product of recombination 
cross section and the concentration of \xm particles. 
It can be easily shown that for 
$Y_X < 0.01$ the recombination rate is too slow to lead to a significant depletion of 
\bes. Detailed calculation of recombination rate and numerical 
analyses of the CBBN at 30 keV \cite{Bird:2007ge,Kusakabe:2007fv} find that the suppression of \bes\
by a factor of 2 is possible for $Y_X \ge 0.1$  if only mechanism (\ref{pburning}) is operative,
and for $Y_X \ge 0.02$ if the
internal conversion (\ref{intconversion}) is allowed. 

{\em Catalysis at 8 keV: enhancement of} \lisx\ {\em and} \ben.

As the Universe continues to cool below 10 keV, an efficient formation of 
(\hef\xm) bound states becomes possible. 
With the reasonable assumption of $Y_{X} < Y_{\rm He}$
the rate of formation of bound states per \xm\ particle is given by the recombination 
cross section and the concentration of the helium nuclei. 
Numerical analysis of recombination reveals that at $T \simeq 5 $ keV about  50\% of available
\xm\ particles will be in bound states with \hef \cite{Pospelov:2006sc}. 

As soon as (\hef\xm) is formed, new reaction channels open up. In particular, 
a photonless thermal production of \lisx\ becomes possible
\bee
(\hefm X^-) + {\rm D} \to \lisxm + X^-;~~ Q \simeq 1.13{\rm MeV} ,
\label{cbbnLi6}
\ee
which exceeds the SBBN production rate by $\sim$six orders of magnitude. The production of
\ben\ may also be catalyzed, possibly by many orders of magnitude relative 
to the SBBN values, 
with the following thermal nuclear chain \cite{Pospelov:2007js}:
\begin{eqnarray}
(\hefm X^-) + \hefm \to (\beetm X^-) + \gamma ;~~ 
(\beetm X^-) + n \to \benm  + X^-.
\label{cbbnBe9}
\end{eqnarray}
Both reactions at these energies are dominated by the 
resonant contributions, although the efficiency of the
second process in (\ref{cbbnBe9}) is not fully understood.

\begin{figure}
\showtwo{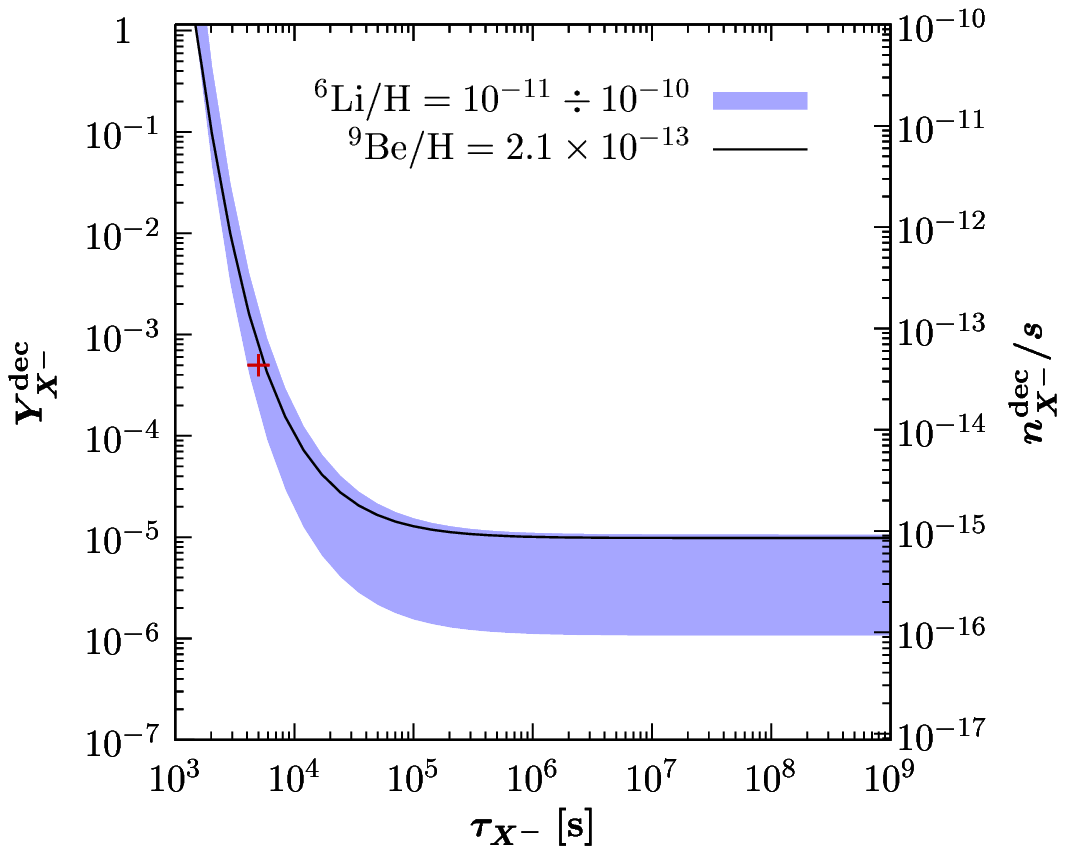}{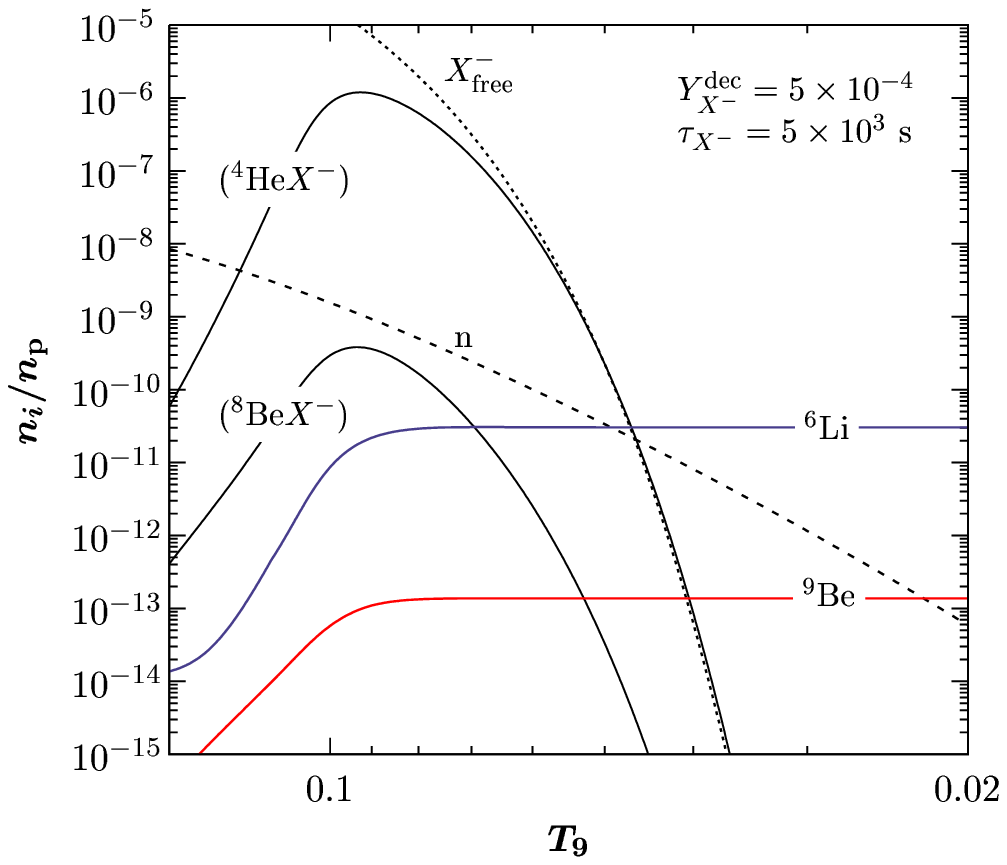}
\caption{Left panel shows CBBN constraints on the abundance vs lifetime of \xm. 
The red cross corresponds to a point in the parameter space, for which the temporal 
development of \lisx\ and \ben\ is shown in the right panel, following Ref.~\cite{Pospelov:2008ta}.
}
\label{fig:BeLi}
\end{figure}

Current estimates/calculations of the CBBN rates are used to determine the generic 
constraints on lifetimes/abundances of charged particles. 
The essence of these limits is displayed in Figure \ref{fig:BeLi},
which shows that for typical \xm\ abundances the lifetime of the 
charged particles would have to be limited by a few thousand seconds! 
This is the main conclusion to be learned from CBBN. Note that while,
to lowest order, 
non-thermal BBN is sensitive to the energy density of decaying particles, 
the CBBN processes are controlled by the number density of \xm, which
underlines the complimentary character of these constraints.
In some models, where both catalysis and cascade nucleosynthesis
occur, catalysis dominates cascade production of \li6 for all
particles with hadronic branching ratio 
$B_h\simle 10^{-2}$~\cite{Jedamzik:2007qk},
whereas \li7 destruction is usually dominated by cascade effects unless
$B_h\simle 10^{-4}$. \be9 production, on the other hand, is conceivable
only through catalysis. 

{ \em Catalysis below 1 keV and nuclear uncertainties}

Finally we comment on the possibility of ($p$\xm) catalysis of nuclear reactions,
discussed in Refs. \cite{Dimopoulos:1989hk,Jedamzik:2007cp}. 
Although it is conceivable that the absence of the Coulomb barrier 
for this compound nucleus may lead to significant changes of 
SBBN/CBBN predictions, 
in practice it turns out that in most cases 
($p$\xm)-related mechanisms are of secondary importance. 
The large radius and shallow binding of this system leads to a 
fast charge-exchange reaction 
on helium,  ($p$\xm) + \hef $\to$ (\hef\xm) + $p$, that reduces the abundance of 
($p$\xm) below $10^{-6}$ relative to hydrogen, 
as long as $Y_{X^-}\simle Y_{\rm ^4He}$, making further 
reactions inconsequential for any observable element \cite{Pospelov:2008ta}. 
In the less likely case, $Y_{X^-}\simge Y_{\rm ^4He}$, significant
late-time processing
due to ($p$\xm) bound states may still occur. Such late time BBN, 
nevertheless, typically leads to observationally 
unacceptable final BBN yields. 

Unlike in the SBBN case and even in cascade nucleosynthesis that utilizes mostly measured nuclear 
reaction rates, CBBN rates cannot be measured in the laboratory, and 
significant nuclear theory input for the calculation of the reaction rates is required. 
However, since the $X^-$ participates only in electromagnetic interactions, 
such calculations are feasible, and dedicated nuclear theory 
studies \cite{Hamaguchi:2007mp} in this direction has already commenced. 
The reaction rates for some CBBN processes, 
such as (\ref{pburning}) and (\ref{cbbnLi6}) are already known 
within a factor of 2 accuracy, and the detailed calculations 
for the \ben\ synthesis are underway \cite{Kamimura:2008fx}.

\section{Dark Matter Production during BBN: NLSP$\to$LSP example}
\label{decay}

\begin{figure}
\begin{center}
\includegraphics[width=.55\textwidth]{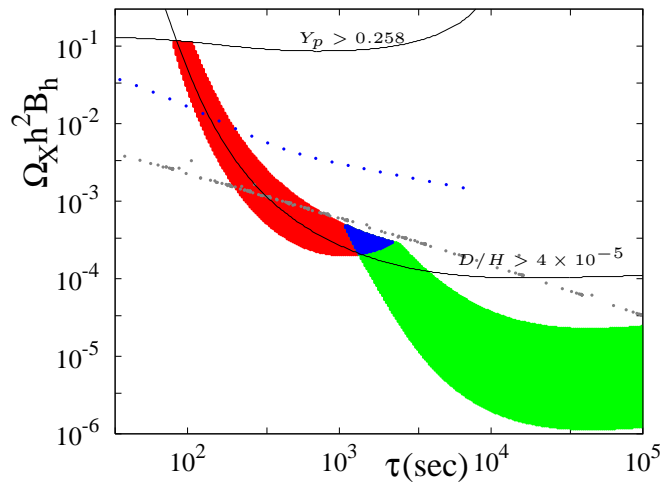}
\end{center}
\caption{Parameter space in the relic decaying particle
abundance times hadronic branching ratio $B_h$, i.e.
$\Omega_Xh^2B_h$, and life time $\tau_X$ plane, where \li7 is significantly
reduced (red and blue) and \li6 is efficiently produced (green and blue).
See text for further details. From Ref.~\cite{Bailly:2008yy}.
}
\label{fig:banana}

\end{figure}

Dark matter particles may be produced by the decay of relic parent particles $X$ during
BBN. Examples, well-studied by different groups, 
include the production of gravitino-LSP dark matter by 
NLSP decays (often charged sleptons or neutralinos) or
production of neutralino dark matter by heavier gravitinos. 
Other conceivable possibilities include the production of superweakly interacting
Kaluza-Klein dark matter, and more generally the cascade decays to any
superweakly interacting dark matter candidates. 
In case of charged NLSP decays, both nonthermal and CBBN processes 
must be accounted for. 
In the framework of gravitino-LSP/stau-NLSP the lifetime of the charged slepton in the 
limit of $m_{\tilde{G}} \ll M_{\rm NLSP}$ is given by
\begin{equation}
\tau_{\rm NLSP}\approx 2.4\times 10^4\,{\rm sec}\,\times
\biggl(\frac{M_{\rm NLSP}}{\rm 300 GeV}\biggr)^{-5}
\biggl(\frac{m_{\tilde{G}}}{\rm 10 GeV}\biggr)^{2}\, ,
\label{eq:lifetime}
\end{equation}
where $M_{\rm NLSP}$ and $m_{\tilde{G}}$ denote NLSP and gravitino mass,
respectively.

\begin{figure}
\begin{center}
\showtwo{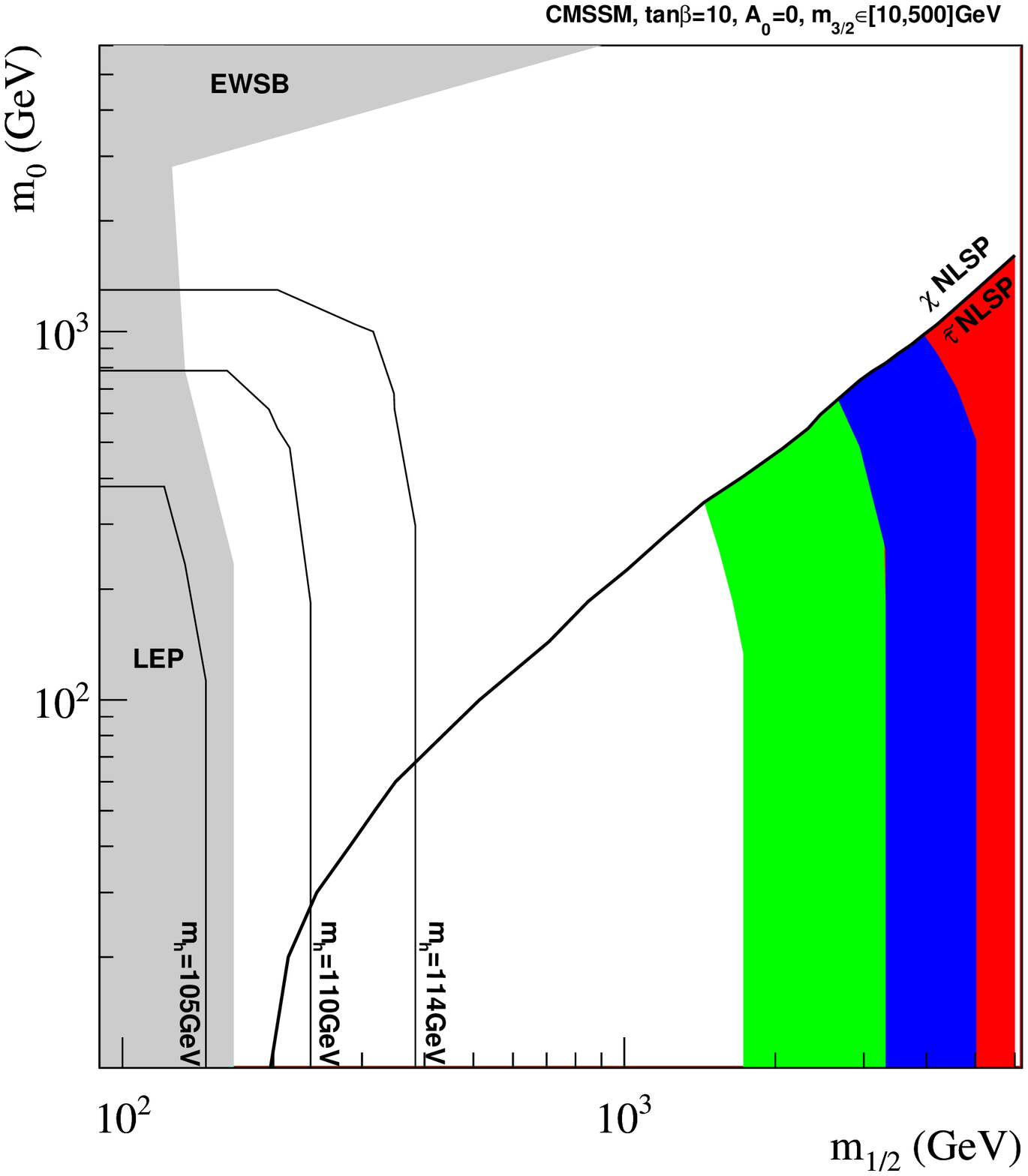}{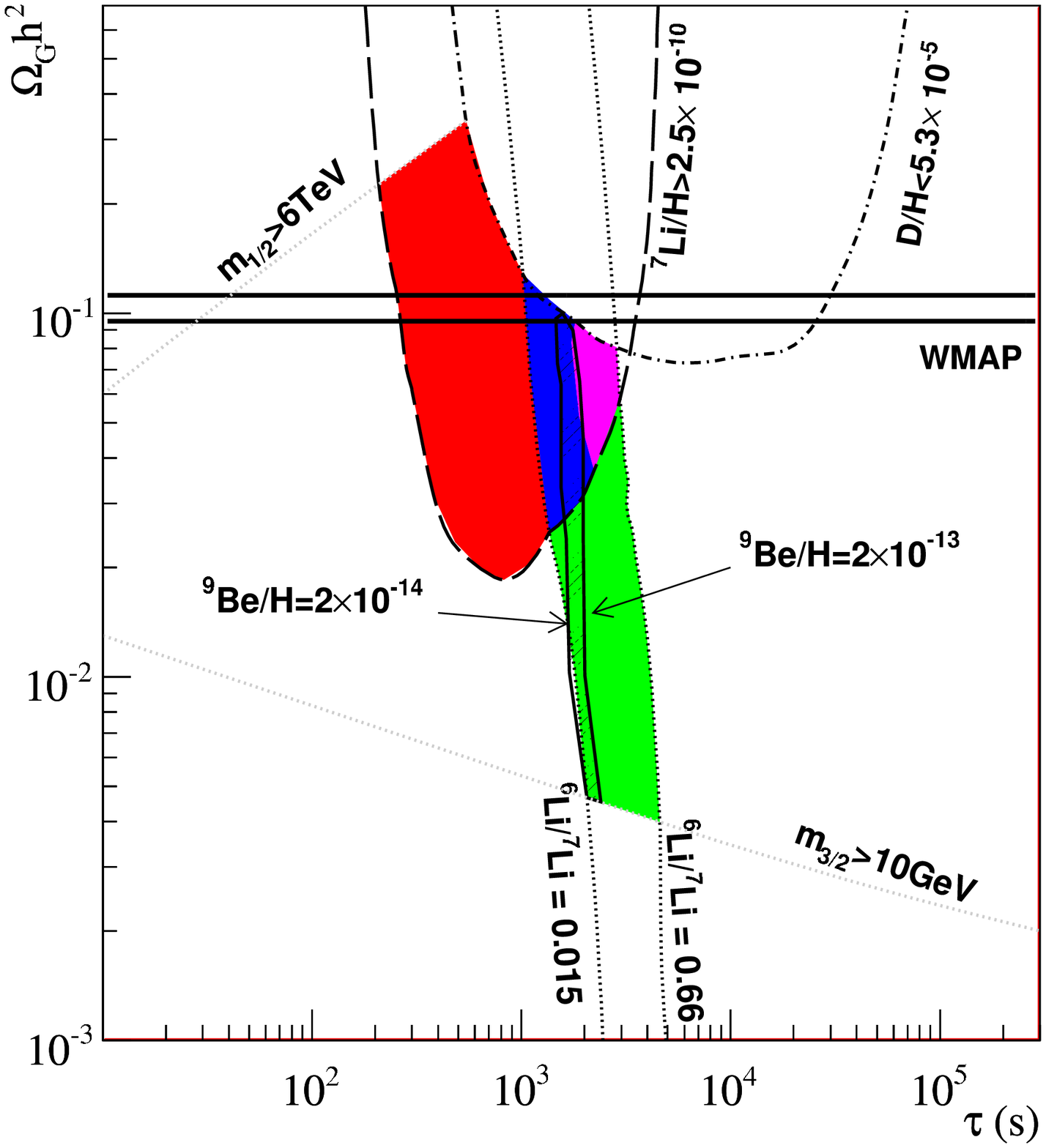}
\end{center}
\caption{Parameter space in the CMSSM which may impact the primordial \li7
and/or \li6 abundances with the color 
coding essentially
as in Fig.~\ref{fig:banana}. Left: unifying scalar soft mass $m_0$
versus gaugino soft mass $m_{1/2}$; Right: gravitino abundance
$\Omega_{\tilde{G}}h^2$ versus NLSP life time $\tau$. 
>From Ref.~\cite{Bailly:2008yy}.
}
\label{fig:CMSSM}
\end{figure}

It becomes exceedingly
more difficult with increasing $\tau_X$ to obtain observational consistency
with inferred primordial abundances (cf. Fig.~\ref{fig:constraint}). 
BBN therefore
plays an important role in 
constraining such scenarios (cf. Sections \ref{cascade} and \ref{catalysis}). 
However, BBN 
may not only constrain, but also favor particular scenarios, 
if current discrepancies with \lisx\ and \lisv\ abundances are to be taken seriously. 
Both trends, the reduction of \lisv\ and the production of \lisx\ via mechanisms described in 
Sections \ref{cascade} and \ref{catalysis}, 
are seen in Fig.~\ref{fig:banana}.
There the red area shows decaying particle
parameter space resulting in more than a factor of 2 suppressed  \li7 abundance 
relative to the 
SBBN prediction, and the green area shows regions where significant
\li6 production ($0.015\simle$\li6/\li7 $\simle 0.3$) occurs. 
In the overlap of these areas, the blue area, both effects may be
achieved simultaneously~\cite{Jedamzik:2004er,Cyburt:2006uv,Cumberbatch:2007me}.
Fig.~\ref{fig:banana} also shows  the prediction of supersymmetric
scenarios with the gravitino-LSP, 
for some representative values of other 
supersymmetric mass parameters. In particular, the grey dots show predictions of stau NLSPs with
gravitino LSPs of mass $m_{\tilde{G}} = 50\,$GeV within the so-called constrained minimal 
supersymmetric SM (CMSSM), 
whereas the blue dots show the case of neutralino NLSPs 
decaying into $m_{\tilde{G}} = 100\,$MeV gravitino LSPs within the  
gauge-mediated supersymmetry breaking scenario.
It is seen that both scenarios naturally cross the region of
"\li7 destruction". The assumption underlying these models is a thermal
freeze-out abundance of the NLSP. 
Since this typically leads to NLSP abundances, 
$10^{-3}\simle\Omega_{\rm NLSP}\simle 10^3$, and taking into account that 
gravitino energy density due to NLSP decays is 
$\Omega_{\tilde{G}} = 
\Omega_{\rm NLSP}\,(m_{\tilde{G}}/M_{\rm NLSP})$,
the resulting $\Omega_{\tilde{G}}$ produced in such scenarios may 
come close to the observed dark matter density. 
This is particularly the case 
for heavy gravitinos $m_{\tilde{G}}\sim 100\,$GeV in the CMSSM, 
for which a more detailed results are shown in Fig.~\ref{fig:CMSSM}.
It is intriguing, and perhaps purely coincidental, 
that when resolving the tension between observed and predicted \lisv\ 
abundance by staus decaying into gravitinos, the resulting
gravitino abundance may account for all the dark matter. For stau decay
times $\tau\approx 10^3$sec it is furthermore possible to synthesize a
primordial \li6 abundance as claimed to be observed in 
low-metallicity stars. Moreover, though less certain, the same parameter 
space could also lead to an important \be9 abundance due to catalytic effects
(cf. Section~\ref{catalysis}), as indicated by the cross-hatched region
in Fig.~\ref{fig:CMSSM}. Finally, since produced by decays, gravitino dark
matter in such scenarios is significantly warm, 
with free-streaming velocities 
of the same order as those
of a $m\approx 3\,$keV early freezing-out relic particle,
which has important implications for the small scale structures in the 
present day Universe. 
%(cf. Section ...).
It is therefore not impossible that some time in the
future, anomalies in the primordial light elements may 
have been understood as signs of the dark matter.
Nevertheless, independent 
verification by particle accelerators, such as 
the LHC is required. Unfortunately, scenarios as
presented in Fig.~\ref{fig:CMSSM} require staus of mass 
$m_{\tilde{\tau}}\simge 1\,$TeV 
too heavy to be produced at the LHC.

\section{Conclusions}

Even though the concept of Big Bang Nucleosynthesis is 
more than 50 yr old, it continues being relevant due to the 
constant progress in nuclear physics, astrophysics, 
and the refined quality of calculations. In this 
paper, we have reviewed the status of BBN, and have shown 
how New Physics can modify the synthesis of light element abundances. 
All three generic ways, extra degrees of freedom modifying the
Hubble expansion during BBN,
energy injection due to annihilation or decay of heavy particles, 
and particle catalysis of BBN reactions,
are highly relevant to the physics associated with particle dark matter,
or with particles intimately tied to dark matter. We have illustrated 
how the existing overall concordance between 
the predicted elemental abundances and observations lead to some 
very non-trivial constraints 
on the properties of DM particles and 
their companions ({\em e.g.} stau-gravitino system). 

Perhaps even more intriguing is the current discrepancy between the 
observed and standard BBN predicted 
abundance of \li7 at the level of 2-3. This discrepancy 
has firmed up since the last unknown SBBN parameter, the 
baryon-to-photon ratio, has been determined with better than 5\% acuracy 
by recent high-precision CMB experiments. At this moment it is 
premature to tell how the \li7 problem is
resolved, but it is nonetheless intriguing that certain models with unstable 
particles 
are capable of alleviating this discrepancy. Hopefully, the continuing 
improvement of 
observational determination of primordial light element abundances, 
as well as the future breakthroughs in electroweak scale particle physics 
would help to solve this important problem.

\vspace{0.5cm}

{\bf References}

\vspace{0.5cm}

\bibliographystyle{iopart-num}
\bibliography{Jedamzik_Pospelov_NJP}

\providecommand{\newblock}{}
\begin{thebibliography}{10}
\expandafter\ifx\csname url\endcsname\relax
  \def\url#1{{\tt #1}}\fi
\expandafter\ifx\csname urlprefix\endcsname\relax\def\urlprefix{URL }\fi
\providecommand{\eprint}[2][]{\url{#2}}

\bibitem{Dunkley:2008ie}
Dunkley J {\em et~al.\/} (WMAP) 2008  (\textit{Preprint} \eprint{0803.0586})

\bibitem{Malaney:1993ah}
Malaney R~A and Mathews G~J 1993 {\em Phys. Rept.\/} {\bf 229} 145--219

\bibitem{Sarkar:1995dd}
Sarkar S 1996 {\em Rept. Prog. Phys.\/} {\bf 59} 1493--1610 (\textit{Preprint}
  \eprint{hep-ph/9602260})

\bibitem{Iocco:2008va}
Iocco F, Mangano G, Miele G, Pisanti O and Serpico P~D 2008  (\textit{Preprint}
  \eprint{0809.0631})

\bibitem{Wagoner:1966pv}
Wagoner R~V, Fowler W~A and Hoyle F 1967 {\em Astrophys. J.\/} {\bf 148} 3--49

\bibitem{Mukhanov:2003xs}
Mukhanov V~F 2004 {\em Int. J. Theor. Phys.\/} {\bf 43} 669--693
  (\textit{Preprint} \eprint{astro-ph/0303073})

\bibitem{Cyburt:2008kw}
Cyburt R~H, Fields B~D and Olive K~A 2008  (\textit{Preprint}
  \eprint{0808.2818})

\bibitem{NaraSingh:2004vj}
Nara~Singh B~S, Hass M, Nir-El Y and Haquin G 2004 {\em Phys. Rev. Lett.\/}
  {\bf 93} 262503 (\textit{Preprint} \eprint{nucl-ex/0407017})

\bibitem{Gyurky:2007qq}
Gyurky G {\em et~al.\/} 2007 {\em Phys. Rev.\/} {\bf C75} 035805
  (\textit{Preprint} \eprint{nucl-ex/0702003})

\bibitem{Brown:2007sj}
Brown T~A~D {\em et~al.\/} 2007 {\em Phys. Rev.\/} {\bf C76} 055801
  (\textit{Preprint} \eprint{0710.1279})

\bibitem{Peimbert:2007vm}
Peimbert M, Luridiana V and Peimbert A 2007  (\textit{Preprint}
  \eprint{astro-ph/0701580})

\bibitem{Izotov:2007ed}
Izotov Y~I, Thuan T~X and Stasinska G 2007 {\em Astrophys. J.\/} {\bf 662}
  15--38 (\textit{Preprint} \eprint{astro-ph/0702072})

\bibitem{Porter:2006gd}
Porter R~L, Ferland G~J and MacAdam K~B 2007 {\em Astrophys. J.\/} {\bf 657}
  327--337 (\textit{Preprint} \eprint{astro-ph/0611579})

\bibitem{Olive:2004kq}
Olive K~A and Skillman E~D 2004 {\em Astrophys. J.\/} {\bf 617} 29
  (\textit{Preprint} \eprint{astro-ph/0405588})

\bibitem{Fukugita:2006}
Fukugita M and Kawasaki M 2006 {\em Astrophys. J.\/} {\bf 646} 691

\bibitem{Burles:1997ez}
Burles S and Tytler D 1998 {\em Astrophys. J.\/} {\bf 499} 699
  (\textit{Preprint} \eprint{astro-ph/9712108})

\bibitem{Levshakov:2001xi}
Levshakov S~A, Dessauges-Zavadsky M, D'Odorico S and Molaro P 2001
  (\textit{Preprint} \eprint{astro-ph/0105529})

\bibitem{Crighton:2004aj}
Crighton N~H~M, Webb J~K, Ortiz-Gill A and Fernandez-Soto A 2004 {\em Mon. Not.
  Roy. Astron. Soc.\/} {\bf 355} 1042 (\textit{Preprint}
  \eprint{astro-ph/0403512})

\bibitem{O'Meara:2006mj}
O'Meara J~M {\em et~al.\/} 2006 {\em Astrophys. J.\/} {\bf 649} L61--L66
  (\textit{Preprint} \eprint{astro-ph/0608302})

\bibitem{Pettini:2008mq}
Pettini M, Zych B~J, Murphy M~T, Lewis A and Steidel C~C 2008
  (\textit{Preprint} \eprint{0805.0594})

\bibitem{Fields:2006ga}
Fields B and Sarkar S 2006  (\textit{Preprint} \eprint{astro-ph/0601514})

\bibitem{Steigman:2007xt}
Steigman G 2007 {\em Ann. Rev. Nucl. Part. Sci.\/} {\bf 57} 463--491
  (\textit{Preprint} \eprint{0712.1100})

\bibitem{Geiss:2007}
Geiss J and Gloeckler G 2007 {\em Space Science Reviews\/} {\bf 130} 5

\bibitem{Sigl:1995kk}
Sigl G, Jedamzik K, Schramm D~N and Berezinsky V~S 1995 {\em Phys. Rev.\/} {\bf
  D52} 6682--6693 (\textit{Preprint} \eprint{astro-ph/9503094})

\bibitem{Ryan:1999jq}
Ryan S~G, Norris J~E and Beers T~C 1999 {\em Astrophys. J.\/} {\bf 523}
  654--677 (\textit{Preprint} \eprint{astro-ph/9903059})

\bibitem{Hosford:2008rs}
Hosford A, Ryan S~G, Perez A~E~G, Norris J~E and Olive K~A 2008
  (\textit{Preprint} \eprint{0811.2506})

\bibitem{Asplund:2005yt}
Asplund M, Lambert D~L, Nissen P~E, Primas F and Smith V~V 2006 {\em Astrophys.
  J.\/} {\bf 644} 229--259 (\textit{Preprint} \eprint{astro-ph/0510636})

\bibitem{Bonifacio:2002yx}
Bonifacio P {\em et~al.\/} 2002  (\textit{Preprint} \eprint{astro-ph/0204332})

\bibitem{Richard:2004pj}
Richard O, Michaud G and Richer J 2005 {\em Astrophys. J.\/} {\bf 619} 538--548
  (\textit{Preprint} \eprint{astro-ph/0409672})

\bibitem{Korn:2006tv}
Korn A {\em et~al.\/} 2006 {\em Nature\/} {\bf 442} 657--659 (\textit{Preprint}
  \eprint{astro-ph/0608201})

\bibitem{Cayrel:1999xx}
Cayrel R, Spite M, Spite F, Vangioni-Flam E, Casse M and Audouze J 1999 {\em
  Astron. and Astrophys.\/} {\bf 343} 923 (\textit{Preprint} \eprint{9901205})

\bibitem{Prantzos:2005mh}
Prantzos N 2005  (\textit{Preprint} \eprint{astro-ph/0510122})

\bibitem{Tatischeff:2006tw}
Tatischeff V and Thibaud J~P 2007 {\em Astron. and Astrophys.\/} {\bf 469} 265
  (\textit{Preprint} \eprint{astro-ph/0610756})

\bibitem{Cayrel:2007te}
Cayrel R {\em et~al.\/} 2007  (\textit{Preprint} \eprint{0708.3819})

\bibitem{Primas:2000gc}
Primas F, Asplund M, Nissen P~E and Hill V 2000  (\textit{Preprint}
  \eprint{astro-ph/0009482})

\bibitem{Boesgaard:2005pf}
Boesgaard A~M and Novicki M~C 2006 {\em Astrophys. J.\/} {\bf 641} 1122--1130
  (\textit{Preprint} \eprint{astro-ph/0512317})

\bibitem{Fields:2004ug}
Fields B~D, Olive K~A and Vangioni-Flam E 2005 {\em Astrophys. J.\/} {\bf 623}
  1083--1091 (\textit{Preprint} \eprint{astro-ph/0411728})

\bibitem{Pospelov:2008ta}
Pospelov M, Pradler J and Steffen F~D 2008 {\em JCAP\/} {\bf 0811} 020
  (\textit{Preprint} \eprint{0807.4287})

\bibitem{Jedamzik:2006xz}
Jedamzik K 2006 {\em Phys.Rev.D\/} {\bf 74} 103509 (\textit{Preprint}
  \eprint{hep-ph/0604251})

\bibitem{Ellis:1984er}
Ellis J~R, Nanopoulos D~V and Sarkar S 1985 {\em Nucl. Phys.\/} {\bf B259} 175

\bibitem{Levitan:1988au}
Levitan Y~L, Sobol I~M, Khlopov M~Y and Chechetkin V~M 1988 {\em Sov. J. Nucl.
  Phys.\/} {\bf 47} 109--115

\bibitem{Dimopoulos:1987fz}
Dimopoulos S, Esmailzadeh R, Hall L~J and Starkman G~D 1988 {\em Astrophys.
  J.\/} {\bf 330} 545

\bibitem{Reno:1987qw}
Reno M~H and Seckel D 1988 {\em Phys. Rev.\/} {\bf D37} 3441

\bibitem{Dimopoulos:1988ue}
Dimopoulos S, Esmailzadeh R, Hall L~J and Starkman G~D 1989 {\em Nucl. Phys.\/}
  {\bf B311} 699

\bibitem{Ellis:1990nb}
Ellis J~R, Gelmini G~B, Lopez J~L, Nanopoulos D~V and Sarkar S 1992 {\em Nucl.
  Phys.\/} {\bf B373} 399--437

\bibitem{Khlopov:1993ye}
Khlopov M~Y, Levitan Y~L, Sedelnikov E~V and Sobol I~M 1994 {\em Phys. Atom.
  Nucl.\/} {\bf 57} 1393--1397

\bibitem{Kawasaki:1994sc}
Kawasaki M and Moroi T 1995 {\em Astrophys. J.\/} {\bf 452} 506
  (\textit{Preprint} \eprint{astro-ph/9412055})

\bibitem{Jedamzik:2004er}
Jedamzik K 2004 {\em Phys. Rev.\/} {\bf D70} 063524 (\textit{Preprint}
  \eprint{astro-ph/0402344})

\bibitem{Kawasaki:2004yh}
Kawasaki M, Kohri K and Moroi T 2005 {\em Phys. Lett.\/} {\bf B625} 7--12
  (\textit{Preprint} \eprint{astro-ph/0402490})

\bibitem{Kawasaki:2004qu}
Kawasaki M, Kohri K and Moroi T 2005 {\em Phys.Rev.D\/} {\bf 71} 083502
  (\textit{Preprint} \eprint{astro-ph/0408426})

\bibitem{Cyburt:2006uv}
Cyburt R~H, Ellis J~R, Fields B~D, Olive K~A and Spanos V~C 2006 {\em JCAP\/}
  {\bf 0611} 014 (\textit{Preprint} \eprint{astro-ph/0608562})

\bibitem{Jedamzik:2004ip}
Jedamzik K 2004 {\em Phys. Rev.\/} {\bf D70} 083510 (\textit{Preprint}
  \eprint{astro-ph/0405583})

\bibitem{Hisano:2003ec}
Hisano J, Matsumoto S and Nojiri M~M 2004 {\em Phys. Rev. Lett.\/} {\bf 92}
  031303 (\textit{Preprint} \eprint{hep-ph/0307216})

\bibitem{Pospelov:2008jd}
Pospelov M and Ritz A 2009 {\em Phys. Lett.\/} {\bf B671} 391--397
  (\textit{Preprint} \eprint{0810.1502})

\bibitem{Adriani:2008zr}
Adriani O {\em et~al.\/} 2008  (\textit{Preprint} \eprint{0810.4995})

\bibitem{Hisano:2008ti}
Hisano J, Kawasaki M, Kohri K and Nakayama K 2009 {\em Phys. Rev.\/} {\bf D79}
  063514 (\textit{Preprint} \eprint{0810.1892})

\bibitem{Pospelov:2006sc}
Pospelov M 2007 {\em Phys. Rev. Lett.\/} {\bf 98} 231301 (\textit{Preprint}
  \eprint{hep-ph/0605215})

\bibitem{Kohri:2006cn}
Kohri K and Takayama F 2007 {\em Phys. Rev.\/} {\bf D76} 063507
  (\textit{Preprint} \eprint{hep-ph/0605243})

\bibitem{Kaplinghat:2006qr}
Kaplinghat M and Rajaraman A 2006 {\em Phys. Rev.\/} {\bf D74} 103004
  (\textit{Preprint} \eprint{astro-ph/0606209})

\bibitem{Hamaguchi:2007mp}
Hamaguchi K, Hatsuda T, Kamimura M, Kino Y and Yanagida T~T 2007 {\em Phys.
  Lett.\/} {\bf B650} 268--274 (\textit{Preprint} \eprint{hep-ph/0702274})

\bibitem{Bird:2007ge}
Bird C, Koopmans K and Pospelov M 2008 {\em Phys. Rev.\/} {\bf D78} 083010
  (\textit{Preprint} \eprint{hep-ph/0703096})

\bibitem{Jittoh:2007fr}
Jittoh T {\em et~al.\/} 2007 {\em Phys. Rev.\/} {\bf D76} 125023
  (\textit{Preprint} \eprint{0704.2914})

\bibitem{Jedamzik:2007cp}
Jedamzik K 2008 {\em Phys. Rev.\/} {\bf D77} 063524 (\textit{Preprint}
  \eprint{0707.2070})

\bibitem{Jedamzik:2007qk}
Jedamzik K 2008 {\em JCAP\/} {\bf 0803} 008 (\textit{Preprint}
  \eprint{0710.5153})

\bibitem{Pospelov:2007js}
Pospelov M 2007  (\textit{Preprint} \eprint{0712.0647})

\bibitem{Kusakabe:2007fv}
Kusakabe M, Kajino T, Boyd R~N, Yoshida T and Mathews G~J 2007
  (\textit{Preprint} \eprint{0711.3858})

\bibitem{Kamimura:2008fx}
Kamimura M, Kino Y and Hiyama E 2008  (\textit{Preprint} \eprint{0809.4772})

\bibitem{DeRujula:1989fe}
De~Rujula A, Glashow S~L and Sarid U 1990 {\em Nucl. Phys.\/} {\bf B333} 173

\bibitem{Dimopoulos:1989hk}
Dimopoulos S, Eichler D, Esmailzadeh R and Starkman G~D 1990 {\em Phys. Rev.\/}
  {\bf D41} 2388

\bibitem{Rafelski:1989pz}
Rafelski J, Sawicki M, Gajda M and Harley D 1991 {\em Phys. Rev.\/} {\bf A44}
  4345

\bibitem{Bailly:2008yy}
Bailly S, Jedamzik K and Moultaka G 2008  (\textit{Preprint}
  \eprint{0812.0788})

\bibitem{Cumberbatch:2007me}
Cumberbatch D {\em et~al.\/} 2007 {\em Phys. Rev.\/} {\bf D76} 123005
  (\textit{Preprint} \eprint{0708.0095})

\end{thebibliography}

\end{document}